\documentclass[prd,twocolumn,aps,showpacs,nofootinbib,nobibnotes,superscriptaddress,preprintnumbers]{revtex4}
\usepackage{epsfig}
\usepackage{graphics}
\usepackage{bm}
\usepackage{color}
\usepackage{dcolumn}   
\usepackage{bm}     
\usepackage{bbm}       
\usepackage{amssymb}  
\usepackage{amsmath}
\usepackage{latexsym}
\usepackage{float}
\usepackage{ifthen}
\usepackage{caption,subfig}
\usepackage{enumerate}
\usepackage{url}
\usepackage{caption,subfig}
\usepackage{amsopn}
\usepackage{hyperref}

\usepackage{amsfonts}
\usepackage{multirow}
\usepackage{array}
\usepackage{booktabs}
\usepackage{rotating}

\usepackage{ulem}
\def\clap#1{\hbox to 0pt{\hss#1\hss}}

\def\({\left(}
\def\){\right)}
\def\[{\left[}
\def\]{\right]}
\def\bea{\begin{eqnarray}}
\def\eea{\end{eqnarray}}
\def\be{\begin{equation}}
\def\ee{\end{equation}}
\def\ba{\begin{eqnarray}}
\def\ea{\end{eqnarray}}
\def\beq{\begin{eqnarray}}
\def\eeq{\end{eqnarray}}

\newcommand{\cs}{c_s}

\newcommand{\mM}{\mathcal{M}}
\newcommand{\Ft}{\tilde{F}}

\newcommand{\Id}{\mathbbm 1}

\def\cs{c_{\rm s}}

\def\be{\begin{equation}}
\def\ee{\end{equation}}
\def\ba{\begin{eqnarray}}
\def\ea{\end{eqnarray}}
\def\beq{\begin{eqnarray}}
\def\eeq{\end{eqnarray}}

\def\L*{{\cal L}_*}
\def\L{\mathcal{L}}
\def\({\left(}
\def\){\right)}

\def\<{\langle}
\def\>{\rangle}

\def\cs2{c_{s}^{2}}

\def\be{\begin{equation}}
\def\ee{\end{equation}}
\def\ba{\begin{eqnarray}}
\def\ea{\end{eqnarray}}
\def\beq{\begin{eqnarray}}
\def\eeq{\end{eqnarray}}

\def\L*{{\cal L}_*}
\def\L{\mathcal{L}}
\def\({\left(}
\def\){\right)}

\def\<{\langle}
\def\>{\rangle}

\begin{document}
%\hspace{5.2in} \mbox{NORDITA-2015-38}\\\vspace{1.53cm} % Preprint number

\title{Derivative self-interactions for a massive vector field}

\date{\today,~ $ $}

\author{Jose Beltr\'an Jim\'enez} \email{jose.beltran@cpt.univ-mrs.fr}
\affiliation{CPT, Aix Marseille Universit\'e, UMR 7332, 13288 Marseille,  France}

\author{Lavinia Heisenberg} \email{lavinia.heisenberg@eth-its.ethz.ch}
\affiliation{Institute for Theoretical Studies, ETH Zurich, 
\\ Clausiusstrasse 47, 8092 Zurich, Switzerland}

\date{\today}

\begin{abstract}
In this work we revisit the construction of theories for a massive vector field with derivative self-interactions such that only the 3 desired polarizations corresponding to a Proca field propagate. We start from the decoupling limit by constructing healthy interactions containing second derivatives of the Stueckelberg field with itself and also with the transverse modes. The resulting interactions can then be straightforwardly generalized beyond the decoupling limit. We then proceed to a systematic construction of the interactions by using the Levi-Civita tensors. Both approaches lead to a finite family of allowed derivative self-interactions for the Proca field. This construction allows us to show that some higher order terms recently introduced as new interactions trivialize in 4 dimensions by virtue of the Cayley-Hamilton theorem. Moreover, we discuss how the resulting derivative interactions can be written in a compact determinantal form, which can also be regarded as a generalization of the Born-Infeld lagrangian for electromagnetism. Finally, we generalize our results for a curved background and give the necessary non-minimal couplings guaranteeing that no additional polarizations propagate even in the presence of gravity.

\end{abstract}

%\pacs{95.35.+d, 04.50.Kd}
%PACS NEEDED

\maketitle

%-----------------------------------------------------------------
\section{Introduction}
 
 \vspace{-0.06cm}
  
The discovery of the cosmic acceleration of the universe triggered a plethora of attempts to unveil the physical mechanism behind it. The simplest explanation comes about in the form of a cosmological constant, but its required small value, although not inconsistent, seriously challenges our theoretical understanding. A natural approach to these somewhat related problems, namely the cosmological constant and the cosmic acceleration, is resorting to infrared (IR) modifications of gravity. Since a gravitational theory based on a massless spin 2 particle needs to coincide with General Relativity (GR) at low energies, modifications of gravity on large distances inevitably lead to the introduction of additional degrees of freedom (dof). In numerous cases, IR modifications of gravity eventually boil down to one additional scalar mode. In the simplest scenarios, it corresponds to a canonical scalar field with a given potential and some couplings to matter. However, in more interesting frameworks, like e.g. the DGP model \cite{Dvali:2000hr}, the additional scalar field gives rise to a novel class of theories characterized by the presence of second order derivative interactions of the scalar field, while the field equations remain of second order, avoiding that way the rise of Ostrogradski instabilities. The properties of this scalar field were then generalized in \cite{Nicolis:2008in} resulting in the class of Galileon theories. These theories are remarkable on their own right because of a number of features, namely: their field equations are explicitly second order even though second derivatives of the fields appear in the action, there is only a finite number of them and are invariant (up to a total derivative) under a constant shift of the field and its gradient, with important consequences for their naturalness under quantum corrections \cite{quantum_corrections}. Interestingly, they have been shown to arise in a natural manner in IR modifications of gravity and played an important role in the construction of a consistent theory of massive gravity \cite{deRham:2010ik,deRham:2010tw}. Moreover, although they modify gravity on large scales, there is a higher scale where new effects come in which is known as Vainshtein radius \cite{Vainshtein:1972sx}. This is in fact a crucial property for the viability of these theories since the scalar field is screened below this scale\footnote{It is worth mentioning that for certain sub-classes of theories, the existence of a Vainshtein screening is not sufficient to avoid conflict with local gravity tests \cite{Babichev:2011iz,Jimenez:2015bwa}.}. The generalization of these theories to include curvature effects led to the (re-)discovery of Horndeski actions as the most general actions for a scalar-tensor theory with second order equations of motion \cite{Horndeski:1974wa}. There exists also an interesting link between massive gravity and these interactions \cite{deRham:2011by}. The Horndeski interactions are however not the most general theories propagating the 2 dof's of the graviton plus 1 additional dof in a scalar-tensor theory \cite{beyondH}.

The construction of Galileon and/or Horndeski actions roots in the same structure found in the Lovelock invariants built by using the symmetry properties of the Levi-Civita tensor and the Bianchi identities. This is actually the reason why the Galileons are typically found in modifications of gravity in higher dimensional setups including Gauss-Bonnet or higher order Lovelock terms \cite{deRham:2010eu}. This line of reasoning was used in \cite{Deffayet:2010zh} to build Galileon-like lagrangians for arbitrary $p$-forms. There it was argued that Galilean interactions are not possible for massless spin 1 fields in 4 dimensions. A more exhaustive classification of Galilean interactions for arbitrary $p$-forms and in arbitrary dimension has been recently performed in \cite{Deffayet:2016von}, where it was confirmed the non-existence of massless vector Galileons in 4 dimensions. This no-go theorem does not extend however to the case of massive spin 1 fields where it is possible to build non-gauge invariant derivative self-interactions of the vector field while keeping the desired 3 propagating degrees of freedom. The key property of these theories is that the Stueckelberg field has the class of Galileon/Horndeski interactions so it only propagates one dof. Interestingly, this type of vector-tensor theories also arise naturally in some modifications of gravity with Gauss-Bonnet terms in Weyl geometries \cite{Jimenez:2014rna,Jimenez:2015fva}.  A classification of derivative vector self-interactions keeping 3 propagating degrees of freedom was carried out in \cite{Heisenberg:2014rta}. A sub-class of these with a coupling of the vector field to the Einstein tensor had been considered in \cite{BeltranJimenez:2010uh} as a potential mechanism to generate cosmic magnetic fields. The case where the longitudinal model has Galilean self-interactions was considered in \cite{Tasinato:2014eka} and its covariantised version in \cite{Heisenberg:2014rta,Tasinato:2014eka,Hull:2015uwa}. Recently, it has been claimed in \cite{Allys:2015sht}  that new derivative self-interactions different from those already found in literature exist and opened the possibility for an infinite series of such terms. This would mean that the massive vector field case is crucially different from its scalar counterpart where Galilean (or, more generally, Horndeski) terms form a finite set of lagrangians. In this note, we revisit this result and argue that the vector field case does resemble the scalar case and a finite series of terms (in a sense that will be made more explicit below) are allowed.

The paper is organized as follows. In the next section we start from the decoupling limit and construct general interactions for the Stueckelberg field containing up to its second derivatives. From this we will then construct theories beyond the decoupling limit. In Section III we will proceed to a systematic construction of the interactions for the massive vector field directly in the unitary gauge by making use of the Levi-Civita tensor. Along with this construction we will show that the higher order derivative self-interactions introduced in  \cite{Allys:2015sht} vanish in 4 dimensions due to a non-trivial cancellation provided by the Cayley-Hamilton theorem. We will then show how the interactions can be nicely rewritten in a determinantal form, which allows to interpret the derivative self-interactions as a generalization of Born-Infeld electromagnetism. Finally, we consider the case of a curved spacetime and give the counter-terms that are needed to avoid additional propagating polarizations when gravity is turned on.

%%%%%%%%%%%%%%%%%%%%%%%%%%%%%%%%
%%%%%%%%%%%%%%%%%%%%%%%%%%%%%%%%
\section{Decoupling limit of generalized Proca}
Historically, the decoupling limit has proven to be advantageous in order to construct healthy theories. Its power lies in its ability to isolate a given degree of freedom and capture its relevant interactions. For instance, in the case of interacting gauge fields, this limit allows to decouple the longitudinal modes together with their self-interactions and study the processes in which they are involved without caring about the remaining transverse modes. The very same 
idea helped with the construction of a non-linear covariant theory of massive gravity without introducing the Boulware-Deser ghost \cite{Boulware:1973my}. In a bottom-up approach, the decoupling limit allowed to isolate the problematic interactions of the helicity-0 mode of the graviton and construct them in a healthy way  \cite{deRham:2010ik}. Once the decoupling limit was under control, it was possible to extend it to a fully non-linear theory. In this section we shall follow an analogous course of action for the case of a Proca field with derivative self-interactions. 

Similarly to the massive gravity case, the non-gauge invariant derivative self-interactions of the vector field might introduce an additional ghostly degree of freedom. In order to be more precise, let us resort to the Stueckelberg trick in order to restore the explicitly broken gauge invariance of a Proca field with mass $M^2$ so that we replace  $A_\mu \to A_\mu+\partial_\mu\pi/M$ with $\pi$ the Stueckelberg field, which will play the role of the longitudinal mode of the massive vector field. If we (carefully) take the limit when the mass goes to zero we can completely decouple $\pi$ and study that sector separately. In the simplest case of a purely massive vector field with $U(1)$ couplings to matter, this limit simply leads to usual electromagnetism with the longitudinal mode being a completely decoupled free massless scalar field. Things are different when considering more general potentials or non-abelian gauge fields, which lead to non-linear sigma models.

 It is the Stueckelberg field which we need to keep under control and make sure that it only propagates the one dof associated to the longitudinal polarization. Since this field does not contribute to the gauge invariant field strength tensor $F_{\mu\nu}=\partial_\mu A_\nu-\partial_\nu A_\mu$, terms built out of $F_{\mu\nu}$ will not introduce the undesired mode. Similarly, since purely potential terms of the form $V(A^2)$ will only introduce first derivatives of the Stueckelberg, they will not add a fourth polarization either. However, when considering non-gauge invariant derivative terms like $(\partial_\mu A^\mu)^2$, the Stueckelberg field will generally acquire higher order derivatives and, thus, an additional mode suffering from the Ostrogradski instability will be present. This pathology can however be bypassed by properly constructing such terms. To that end, we will require the following conditions:
\begin{itemize}
\item The pure Stuckelberg field sector belongs to the Galileon/Horndeski class of lagrangians. Due to the origin of $\pi$, only the subclass with shift symmetry can be present.
\item The couplings of second derivatives of $\pi$ to the transverse modes must also lead to second order field equations.
\end{itemize}
The first condition will be relevant for the leading order in the decoupling limit with interactions purely constructed out of the Stueckelberg field. The second condition will be important for the terms with non-trivial couplings between the transverse modes and the Stueckelberg field. More explicitly, we will consider lagrangians depending on the vector field $A_\mu$ and its first derivatives $\partial_\mu A_\nu$. Since we want to explicitly separate the derivative interactions with non-trivial contributions for $\pi$, we will express the lagrangian as $\L=\L(A_\mu,F_{\mu\nu},S_{\mu\nu})$ with $S_{\mu\nu}=\partial_\mu A_\nu+\partial_\nu A_\mu$. Moreover, we will introduce a given scaling for each object so that the corrections with respect to the pure Proca action admit an expansion of the form
\be
\L\sim\sum_{m,n,p}c_{m,n,p}\left(\frac{A}{\Lambda_M}\right)^m\left(\frac{F}{\Lambda_F^2}\right)^n\left(\frac{S}{\Lambda_S^2}\right)^p \,,
\ee
where $c_{m,n,p}$ are some coefficients and $\Lambda_M$, $\Lambda_F$ and $\Lambda_S$ are some scales suppressing the dependence on each object. In the decoupling limit where the mass $M^2$ is appropriately sent to zero 
the vector field will be dominated by its longitudinal polarization or, in other words, $A_\mu$ essentially becomes $\partial_\mu\pi/M$. This further implies that $S_{\mu\nu}$ will become $\partial_\mu\partial_\nu\pi/M$ in that limit. Finally, since $F_{\mu\nu}$ is gauge invariant, the Stueckelberg field will not contribute to it. Thus, in this limit, the lagrangian will take the schematic form
\be
\L_{\rm dec}\sim\sum_{m,n,p}c_{m,n,p}\left(\frac{\partial\pi}{M\Lambda_M}\right)^m\left(\frac{F}{\Lambda_F^2}\right)^n\left(\frac{\partial\partial\pi}{M\Lambda_S^2}\right)^p \,.
\label{Ldec}
\ee
Now we can focus at each order in the second derivatives of $\pi$ in order to build the healthy interactions that will prevent the propagation of an additional mode for the Stueckelberg field. Moreover, at each order the required properties and tensor structure will allow us to perform the resummations in $m$ and $n$. By looking at the structure in \ref{Ldec} we realize that the problem reduces to finding healthy second derivative interactions for $\pi$ with itself and with a gauge field. The self-interactions have already been extensively analysed in the literature and are given by the well-known Galileon/Horndeski types of terms. The couplings to gauge fields is more delicate, but they have also been considered in the literature\footnote{A related topic is the case when Galileon fields have some gauge symmetry, in which case consistent couplings between the Galileon fields and gauge fields also arise \cite{Zhou:2011ix}.} (see e.g. \cite{Heisenberg:2014rta,Tasinato:2013oja}). We shall now proceed to the construction of the interactions order by order in the second derivatives of $\pi$. Let us start by the lowest order with $p=1$, i.e, linear in $\partial\partial\pi$.  At leading order the healthy interactions are given by
\be
\L\sim\left(c_{2,0,1}(\partial\pi)^2\eta^{\mu\nu}+c_{0,2,1}\Ft^{\mu\alpha}\Ft^\nu{}_\alpha\right)\frac{\partial_\mu\partial_\nu\pi}{M\Lambda_S^2} \,,
\label{Ldecp1}
\ee
where $\Ft^{\mu\nu}\equiv \frac12 \epsilon^{\mu\nu\alpha\beta}F_{\alpha\beta}$ is the dual of the strength tensor. The first term is just the usual cubic Galileon while the second term arises as the generalization of Galileon interactions for mixed 0- and 1-forms. Notice that a possible term proportional to $\partial^\mu\pi\partial^\nu\pi$ is equivalent to the cubic Galileon via integration by parts. Now we can straightforwardly proceed to a partial resummation in $m$ and $n$. The final term will take the form $\mM^{\mu\nu}\partial_\mu \partial_\nu\pi$ where $\mM^{\mu\nu}$ is some symmetric rank-2 tensor built out of the metric $\eta^{\mu\nu}$, $\partial_\mu\pi$ and $F_{\mu\nu}$. In general, this type of terms in the action guarantees the absence of higher than second time derivatives field equations if $\mM^{00}$ does not contain time derivatives others than $\dot{\pi}$. This is precisely what occurs in (\ref{Ldecp1}) since $\Ft^{0\alpha}\Ft^0{}_\alpha\propto B^2$, with $B$ the magnetic part of $F_{\mu\nu}$. Since the magnetic part is purely potential (it does not contain time derivatives of $A$), the structure in (\ref{Ldecp1}) guarantees no higher than second derivative field equations. It turns out that this is the only coupling of this form that satisfies this requirement. Thus, we can resum by simply promoting the coefficients $c_{m,n,1}$ into arbitrary functions of $X\equiv\partial_\mu \pi \partial^\mu\pi$. Moreover, the metric tensor can also be promoted into a disformal metric of the form $\eta^{\mu\nu}+g(x)\partial^\mu\pi\partial^\nu\pi$ without spoiling the healthy properties of the decoupling limit. Thus, we end up with
\begin{align}
\L&\sim \Big[c_{1}(X)\eta^{\mu\nu}+g_1(X)\partial^\mu\pi\partial^\nu\pi\nonumber\\
&+c_{2}(X)\Big(\eta^{\alpha\beta}+g_2(X)\partial^\alpha\pi\partial^\beta\pi\Big)\Ft^\mu{}_\alpha\Ft^\nu{}_\beta\Big]\frac{\partial_\mu\partial_\nu\pi}{M\Lambda_S^2}.
\label{Lfullp1}
\end{align}
Notice that the second order nature of the field equations for the transverse modes is guaranteed by the Bianchi identities making $\Ft^{\alpha\beta}$ divergence-free. Also, as commented above, the second term in the first line is equivalent to the first term up to integration by parts. However, when going beyond the decoupling limit, this term will result in non-trivial interactions for the vector field, so it is convenient to keep it explicitly.

Now let us turn to the terms quadratic in $\partial\partial\pi$ corresponding to $p=2$. In this case, the possible interactions are
\begin{align}
\L\sim &c_{2,0,2}(\partial\pi)^2\frac{(\Box\pi)^2-(\partial_\mu\partial_\nu\pi)^2}{M^2\Lambda_S^4}\nonumber\\&
+c_{0,2,2}\Ft^{\mu\nu}\Ft^{\alpha\beta}\frac{\partial_\mu\partial_\alpha\pi\partial_\nu\partial_\beta\pi}{M^2\Lambda_S^4} \,,
\label{Ldecp2}
\end{align}
where again we recognize the typical quartic Galileon interactions in the first line, while in the second line we have the mixing with the gauge field. Again, we can check that second time derivatives of $\pi$ only couple to the magnetic part of the gauge field, and this allows to avoid higher order equations of motion for $\pi$. On the other hand, since $\Ft^{\mu\nu}$ is divergence-free, the gauge field equations will also remain of second order. Analogously to the previous case, the tensorial structure that we have in (\ref{Ldecp2}) will persist in a partial resummation on $m$ and $n$ so we can promote the coefficients to arbitrary functions of $X$, along with the aforementioned disformally transformed metric.

The next order corresponds to $p=3$ and the lagrangian reduces to the quintic Galileon interaction for $\pi$. At this order it is not possible to construct healthy mixed interactions with the gauge field and the same applies for higher $p>3$ orders. Thus, from that order on no new interactions are possible (notice that the Galileon terms also stop at this order) and the series stops at $p=3$ \footnote{Let us mention that this is true in 4 dimensions, but other interactions are possible in higher dimensions $d>4$. Besides the usual Galileon interactions, the mixing with the gauge field can be generalized to 
\be
\Ft^{\mu_1\cdots\mu_{n}}\Ft^{\nu_1\cdots\nu_{n}}\partial_{\mu_1}\partial_{\nu_1}\pi\cdots\partial_{\mu_i}\partial_{\nu_i}\pi\eta_{\mu_{i+1}\nu_{i+1}}\cdots\eta_{\mu_{n}\nu_{n}}
\label{higherdimensions}
\ee
with $n=d-2$.}.
 From the resulting lagrangians in the decoupling limit it is now straightforward to construct the full theory by simply replacing $\partial_\mu\pi\rightarrow A_\mu$ and $\partial_\mu\partial_\nu\pi\rightarrow S_{\mu\nu}$. 

After arguing how to generate the lagrangians for the generalized Proca theories from the decoupling limit we will now turn to the general case and proceed to the construction of the different terms in a systematic way directly beyond the decoupling limit.

%%%%%%%%%%%%%%%%%%%%%%%%%%%%%%%%
%%%%%%%%%%%%%%%%%%%%%%%%%%%%%%%%
\section{Systematic construction}
In this Section we will resort to the useful antisymmetric properties of the Levi-Civita tensor in order to build the derivative self-interactions for the generalized Proca action. This will allow us to recover the interactions obtained in the previous section directly for the full action without resorting to the decoupling limit. As we argued above within the decoupling limit, we expect the derivative self-interactions to form a finite series and the construction of this section will also point towards the same conclusion. As in the previous Section, we shall proceed order by order, in derivatives of $A_\mu$ this time. Starting with the interactions linear in $\partial A$ we can have
 \begin{equation}\label{GenProcaL3}
\mathcal{L}_3=-\frac{f_3(A^2)}{6}\epsilon^{\mu\nu\rho\sigma}\epsilon^{\alpha}_{\;\;\nu\rho\sigma}\partial_\mu A_\alpha=\frac{f_3}{2}[S] \,,
 \end{equation}
where $[\cdots]$ stands for the trace (with respect to the Minkowski metric) of the matrix. This term leads to a cubic Galileon interaction for the longitudinal polarization in the decoupling limit. At this stage there is only one way of contracting the indices with the antisymmetric tensor since $\epsilon^{\mu\nu\rho\sigma}\epsilon^{\alpha}_{\;\;\nu\rho\sigma}\partial_\alpha A_\mu$ would give rise exactly to the same interaction $(\partial\cdot A)$. However, for the sake of generality, we should emphasize that the hidden metrics in (\ref{GenProcaL3}) used to contract the indices of the Levi-Civita tensors among themselves can always be replaced by a disformally transformed metric analogously to the case discussed for the decoupling limit, i.e., we can always replace $\eta_{\mu\nu}\rightarrow \eta_{\mu\nu}+g(A^2)A_\mu A_\nu$. This will equivalently generate for instance interactions of the form $f_3A^\mu A^\nu (\partial_\mu A_\nu)$ at this order. This interaction is set on equal footing as the previous one and, in fact, it is equivalent to it since $f_3A^\mu A^\nu \partial_\mu A_\nu=\frac12 f_3\partial_\mu A^2A^\mu=\partial_\mu F_3 A^\mu$ with $F_3'=f_3$.

As next we shall consider terms quadratic in derivatives of the vector field, i.e., containing $(\partial A)^2$. For these terms we can build two different ways of contracting the Levi-Civita indices, namely
\begin{eqnarray}
\mathcal{L}_4&=&-\frac{f_4(A^2)}{2}\epsilon^{\mu\nu\rho\sigma}\epsilon^{\alpha\beta}_{\;\;\;\;\rho\sigma}(\partial_\mu A_\alpha\partial_\nu A_\beta+c_2\partial_\mu A_\nu\partial_\alpha A_\beta)\nonumber\\
&=&\frac{f _4}{4}\Big([S]^2-[S^2]+(1+2c_2)[F^2]\Big) \,.
\end{eqnarray}
The term proportional to $c_2$ just renormalizes the standard kinetic term as it corresponds to $c_2F_{\mu\nu}^2$ and hence does not contain the dependence on the longitudinal mode. Of course we could have chosen yet another function in front of the $c_2$ term instead of $f_4$ but we are discarding this contribution anyway since we will include a kinetic term for the vector field separately. Note again, that if we had contracted the indices of the Levi-Civita tensors with the vector field $A_\mu$ instead of the metric, we would have generated terms of the form $f _4A^\mu A^\nu(\partial_\nu A_\mu(\partial\cdot A)-\partial_\nu A_\rho \partial^\rho A_\mu)$, which are again at the same footing as the $\mathcal{L}_4$ interaction we constructed above.  The quintic interactions can be constructed in a similar way. There will be again two different ways of contracting the indices
\begin{eqnarray}
\mathcal{L}_5&=&-f_5(A^2)\epsilon^{\mu\nu\rho\sigma}\epsilon^{\alpha\beta\delta}_{\;\;\;\;\;\;\sigma}\partial_\mu A_\alpha\partial_\nu A_\beta \partial_\rho A_\delta \nonumber\\
&-&d_2f_5(A^2)\epsilon^{\mu\nu\rho\sigma}\epsilon^{\alpha\beta\delta}_{\;\;\;\;\;\;\sigma}\partial_\mu A_\nu\partial_\rho A_\alpha \partial_\beta A_\delta \nonumber\\
& =&\frac{f_5}8\Big([S]^3-3[S][S^2]+2[S^3] \nonumber\\
 &&-2(3+2 d_2)\Ft^{\mu\alpha}\Ft^\nu{}_\alpha S_{\mu\nu}\Big)\,.
\end{eqnarray}
Again, we could have chosen a different function in front of the $d_2$ term instead of $f_5$ but we have chosen it to be the same for now just for compactness of the expression. If we had imposed the condition that the scalar part of the vector field should only have terms that do not correspond to any trivial total derivative interactions, then the series would stop here \cite{Heisenberg:2014rta}. In fact if we relax this condition, we can construct yet another order of interactions. These sixth order interactions have also two different ways of contracting the indices\footnote{The terms purely depending on $S$ are just a total derivative and do not contribute to the equation of motion. This is consistent with the fact that the scalar Galileons at that order have purely total derivative interactions and the series of the scalar Galileons hence stops at $\mathcal{L}_5$.}
\begin{align}
\mathcal{L}_6=&-f_6(A^2)\epsilon^{\mu\nu\rho\sigma}\epsilon^{\alpha\beta\delta\kappa}\partial_\mu A_\alpha\partial_\nu A_\beta \partial_\rho A_\delta \partial_\sigma A_\kappa \nonumber\\
&-e_2f_6(A^2)\epsilon^{\mu\nu\rho\sigma}\epsilon^{\alpha\beta\delta\kappa}\partial_\mu A_\nu\partial_\alpha A_\beta \partial_\rho A_\delta  \partial_\sigma A_\kappa\nonumber\\
&=\frac{f_6}{16}\Big((3+2e_2)([F^2]^2-2[F^4]) \nonumber\\
&-4(3+e_2)\Ft^{\mu\nu}\Ft^{\alpha\beta}S_{\mu\alpha}S_{\nu\beta} +[S]^4 \nonumber\\
&-6[S]^2[S^2]+3[S^2]^2+8[S][S^3]-6[S^4]\Big)
\end{align}
The terms purely depending on $F$ will not contribute to the longitudinal mode and we will group them into $\mathcal L_2$. Note that we have not included the contraction $\epsilon^{\mu\nu\rho\sigma}\epsilon^{\alpha\beta\delta\kappa}\partial_\mu A_\nu\partial_\alpha A_\beta \partial_\rho A_\sigma  \partial_\delta A_\kappa$ either, since this will also give purely gauge invariant quantities. The sixth order interactions are in agreement with \cite{Allys:2015sht}, just written in a slightly different way. The total Lagrangian for the vector field is
\begin{equation}\label{generalizedProfaField}
\mathcal L_{\rm gen. Proca} = -\frac14 F_{\mu\nu}^2 +\sum^5_{n=2}\alpha_n \mathcal L_n \,,
\end{equation}
where the self-interactions of the vector field are
\begin{eqnarray}\label{vecGalProcaField}
\mathcal L_2 & = &f_2(A_\mu, F_{\mu\nu}, \tilde{F}_{\mu\nu})\nonumber\\
\mathcal L_3 & = &f_3(A^2) \;\; \partial\cdot A \nonumber\\
\mathcal L_4  &=&  f _4(A^2)\;\left[(\partial\cdot A)^2-\partial_\rho A_\sigma \partial^\sigma A^\rho\right]   \nonumber\\
\mathcal L_5  &=&f_5(A^2)\;\left[(\partial\cdot A)^3-3(\partial\cdot A)\partial_\rho A_\sigma \partial^\sigma A^\rho \right. \nonumber\\
&&\left.+2\partial_\rho A_\sigma \partial^\gamma A^\rho\partial^\sigma A_\gamma \right]  +\tilde{f}_5(A^2)\tilde{F}^{\alpha\mu}\tilde{F}^\beta_{\;\;\mu}\partial_\alpha A_\beta \nonumber\\
\mathcal L_6  &=&f_6(A^2) \tilde{F}^{\alpha\beta}\tilde{F}^{\mu\nu}\partial_\alpha A_\mu \partial_\beta A_\nu \,.
\end{eqnarray}
Note that the series stops here and there are not any higher order terms beyond the sixth order interactions. An interesting question is whether one can take functions of these invariants and build new terms in the similar spirit as $f(R)$-theories. Any combinations of these invariants however are expected to propagate at least one more degree of freedom.
%%%%%%%%%%%%%%%%%%%%%%%%%%%%%%%%
%%%%%%%%%%%%%%%%%%%%%%%%%%%%%%%%

Within the framework of the systematic construction of the generalized Proca interactions in terms of the Levi-Civita tensors (also supporting our findings in the decoupling limit), we observe that the series stops after the sixth order of interactions. In other words, there are no indices left in the two Levi-Civita tensors in order to construct a possible $\mathcal L_7$ term in 4 dimensions. In \cite{Allys:2015sht} it was argued that the following higher order term at seventh order exist
\begin{align}\label{L7Perm1}
&\mathcal L_7^{\rm Perm,1}=(\partial\cdot A)^3F_{\mu\nu}^2+6(\partial\cdot A)^2\partial^\mu A^\nu \partial^\rho A_\mu F_{\nu\rho} \nonumber\\
&+3(\partial\cdot A)((\partial_\nu A_\rho \partial^\rho A^\nu)^2-(\partial_\nu A_\rho \partial^\nu A^\rho)^2) \nonumber\\
&+3(\partial\cdot A)\partial^\mu A_\nu (F^\nu{}_\rho F^\rho{}_\sigma F^\sigma{}_\mu -4  \partial^\nu A_\rho \partial^\rho A_\sigma F^\sigma_\mu) \nonumber\\
&+4\partial_\mu A_\nu (\partial^\mu A^\nu \partial^\rho A^\sigma \partial_\gamma A_\rho \partial^\gamma A_\sigma - \partial^\nu A^\mu \partial^\rho A^\sigma \partial_\gamma A_\rho \partial_\sigma A^\gamma) \nonumber\\
&+2\partial_\mu A_\nu \partial^\mu A^\nu \partial^\rho A_\sigma\partial_\gamma A_\rho F^{\gamma\sigma}-6 \partial^\mu A_\nu F^\nu_{\;\;\rho} F^\rho_{\;\;\sigma} F^\sigma_{\;\;\gamma}F^\gamma_{\;\;\mu}\nonumber\\&+12 \partial^\mu A_\nu \partial^\nu A_{\rho}\partial^\rho A_{\sigma}\partial^\sigma A_{\gamma}F^\gamma{}_{\mu} \,,
\end{align}
and similarly
\begin{align}\label{L7Perm2}
\mathcal L_7^{\rm Perm,2}=\frac{1}4 (\partial\cdot A)((F_{\mu\nu}F^{\mu\nu})^2-4\partial^\mu A_\nu  F^\nu_{\;\;\rho} F^\rho_{\;\;\sigma} F^\sigma_{\;\;\mu}) \nonumber\\
+(F_{\mu\nu}F^{\mu\nu})\partial^\sigma A^\rho \partial^\gamma A_\sigma F_{\rho\gamma}+2\partial^\mu A_\nu F^\nu_{\;\;\rho} F^\rho_{\;\;\sigma} F^\sigma_{\;\;\gamma}F^\gamma_{\;\;\mu}\,.
\end{align}
It was argued that these two seventh order derivative interactions give rise to interactions which propagate three degrees of freedom and hence extend the vector Galileons beyond the previous order. Based on this they conjectured that the vector Galileons have infinite series of such interactions. These terms were found by imposing the vanishing of the determinant of the Hessian matrix $H^{\mu\nu}=\partial^2 \L/\partial \dot{A}_\mu\partial \dot{A_\nu}$ so that it is guaranteed the existence of one primary constraint that will remove the undesired polarization for the vector field. In particular, they looked for lagrangians satisfying  $H^{0\mu}=0$ aiming at obtaining a constraint for the temporal component. Solving for this condition the above interactions were obtained. However, once those terms are obtained their full constraints structure should be carefully checked in order to guarantee that the constraint is actually second class. Were it be first class it would generate a gauge symmetry that would remove an additional polarization. Similarly, it could happen that the obtained interactions could reduce to total derivatives or even trivialize. We have explicitly checked these requirements for the higher order terms (\ref{L7Perm1}) and (\ref{L7Perm2}) that were proposed in \cite{Allys:2015sht} and found that they correspond to trivial interactions in the sense that they vanish exactly. However, the vanishing of these terms is due to non-trivial relations. To show it we will take (\ref{L7Perm2}) and express everything in terms of $F_{\mu\nu}$ and $S_{\mu\nu}$ after which we obtain $\mathcal L_7^{\rm Perm,2}=M^\mu{} _{\nu} S_\mu{}^\nu$ with
\be
M=F^4-\frac12 [F^2]F^2+\frac18\Big([F^2]^2-2[F^4]\Big)\Id
\label{CHtheoremL7}
\ee
where we have used matrix notation and $\Id$ stands for the identity matrix. We can recognize that $M$ vanishes in 4 dimensions by virtue of the Cayley-Hamilton theorem applied on $F^\mu{}_\nu$ and, consequently, the interaction is trivial in 4 dimensions, but it can be present in higher dimensions. We have checked this explicitly for both (\ref{L7Perm1}) and (\ref{L7Perm2}) and found that they are healthy interactions in 5 dimensions, but they trivialize in 4 dimensions. Thus, the seventh order terms only give rise to new interactions in dimensions higher than 4. It is worth reminding here that in higher dimensions we expect new interactions very much like in the scalar Galileons case (see also footnote 3 for the construction of higher order interactions in the vector Galileon case in higher dimensions). Our statement can be easily shown by noticing that the interaction $\mathcal L_7^{\rm Perm,2}$ can be written as
\begin{align}
\mathcal L_7^{\rm Perm,2}=-\frac{1}{2}\epsilon^{\mu\nu\rho\sigma\tau}\epsilon^{\alpha\beta\delta\kappa\omega}\partial_\mu A_\nu\partial_\alpha A_\beta \partial_\rho A_\sigma  \partial_\delta A_\kappa  \partial_\tau A_\omega\; .
\end{align}
Since we need the Levi-Civita symbol with 5 indices, this term identically vanishes in 4 dimensions. This is nothing but an alternative way of writing the Cayley-Hamilton theorem. In fact, we can express the $\epsilon$'s in terms of metric tensors to re-obtain (\ref{CHtheoremL7}). The same applies to $\mathcal L_7^{\rm Perm,1}$ since that term can be expressed as
\begin{align}
\mathcal L_7^{\rm Perm,1}=&-3\mathcal L_7^{\rm Perm,2}\\
&-2\epsilon^{\mu\nu\rho\sigma\tau}\epsilon^{\alpha\beta\delta\kappa\omega}\partial_\mu A_\nu\partial_\alpha A_\beta \partial_\rho A_\delta \partial_\sigma A_\kappa \partial_\tau A_\omega \,.\nonumber
\end{align}
Again, we need at least 5 dimensions for this term not to vanish. For completeness, we can also give here the purely Galileon interaction which is only non-trivial in dimensions higher than 4:
\begin{align}
\mathcal{L}_7^{\rm Gal}=\epsilon^{\mu\nu\rho\sigma\tau}\epsilon^{\alpha\beta\delta\kappa\omega}\partial_\mu A_\alpha\partial_\nu A_\beta \partial_\rho A_\delta \partial_\sigma A_\kappa \partial_\tau A_\omega \,.
\end{align}
We can mention that, in fact, all the above terms will generate interactions of the form given in (\ref{higherdimensions}) for higher dimensions, showing that they naturally follow from both our analysis in the decoupling limit and our systematic construction.

We shall end this Section by noticing that the structure of the interactions based on the antisymmetry of the Levi-Civita tensor allows a nice determinantal formulation of the generalized Proca interactions. The existence of such a formulation is not surprising and it is in the same spirit as the interactions in massive gravity and scalar Galileon interactions, which can also be compactly written in terms of a determinantal interaction. In our case, the generating determinant can be expressed as
\begin{equation}\label{determinant}
f(A^2)\det(\delta^{\mu}{}_\nu+\mathcal{C}^\mu{}_\nu)
\end{equation}
with the fundamental matrix 
\be
\mathcal{C}^\mu{}_\nu=aF^\mu{}_\nu +bS^\mu{}_\nu +c A^\mu A_\nu
\ee
with $a$, $b$ and $c$ some parameters of dimension $-2$. We have also included the arbitrary function $f(A^2)$ for completeness. Notice that this is the more general matrix that can be built with the vector field and up to its first derivatives at the lowest order. Moreover, we can easily make contact with the decoupling limit by assuming that the constants $a$, $b$ and $c$ scale as $\Lambda_F^{-2}$, $\Lambda_S^{-2}$ and $\Lambda_M^{-2}$ respectively. A special case is the one with $a=-b$ and $c=0$, since in that case the determinant reduces to $\det(\delta^{\mu}{}_\nu+2b\partial_\nu A^\mu)$, which can be identified with the Jacobian of a coordinate transformation $x^\mu\rightarrow x^\mu+2b A^\mu$. If we take $f(A^2)=A^2$, this is precisely what one obtains from the pure Proca mass term after applying a generalised Galileon transformation \cite{deRham:2014lqa}.

From the above determinant, the derivative self-interactions for the vector field can be easily implemented via the relation 
\be
\det(\delta^{\mu}{}_\nu+\mathcal{C}^\mu{}_\nu)=\sum_{n=0}^4 e_n(\mathcal{C}^\mu{}_\nu)
\ee
with $e_n$ the elementary symmetric polynomials of the fundamental matrix $\mathcal{C}^\mu{}_\nu$ whose expressions are given below. The zeroth order symmetric elementary polynomial is trivial $e_0 = 1$ whereas the first order yields simply
\begin{eqnarray}
e_1 &=&-\frac{1}{6}\epsilon_{\mu\nu\alpha\beta}\epsilon^{\rho\nu\alpha\beta}\mathcal{C}^\mu{}_\rho=[\mathcal{C}]
= b[S]+cA^2\,.
\end{eqnarray}
The first term is nothing else but $\mathcal L_3$ and the second term just a potential interaction as a part of $\mathcal L_2$. The first interesting vector Galileon interaction is encoded in the second order symmetric elementary polynomial
\begin{eqnarray}
e_2 &=&-\frac{1}{4}\epsilon_{\mu\nu\alpha\beta}\epsilon^{\rho\sigma\alpha\beta}\mathcal{C}^\mu{}_\rho\mathcal{C}^\nu{}_\sigma=\frac{1}{2}\Big([\mathcal{C}]^2-[\mathcal{C}^2]\Big) \nonumber\\
&=&\frac{1}{2}\Big( a^2[F^2]+b^2([S]^2-[S^2])\nonumber\\
&+&  2bc(A^2\eta^{\mu\nu}-A^\mu A^\nu )S_{\mu\nu} \Big). 
\end{eqnarray}
The terms of the first line of this equation correspond to the vector interactions that we constructed explicitly in $\mathcal L_4$ and the first two terms in the second line are simply of the type $\mathcal L_3$ when we contracted the indices of the Levi-Civita tensors with the vector fields instead of the metrics. Similarly, we can construct the cubic symmetric elementary polynomial
\begin{eqnarray}
e_3 &=&-\frac{1}{6}\epsilon_{\mu\nu\alpha\beta}\epsilon^{\rho\sigma\delta\beta}\mathcal{C}^\mu{}_\rho\mathcal{C}^\nu{}_\sigma \mathcal{C}^\alpha{}_\delta   \nonumber\\
&=& \frac{a^2c}{2} F_\alpha{}^\mu F_{\beta\mu}(A^2 \eta^{\alpha\beta}-2A^\alpha A^\beta ) \nonumber\\
&+&\frac{1}2 a^2b ([F^2][S]-2F^{\alpha\beta}F_{\alpha}{}^\mu S_{\beta\mu})  \nonumber\\
&+&\frac12cb^2 S_\alpha{}^\mu S_{\beta\mu}(2A^\alpha A^\beta -A^2\eta^{\alpha\beta})  \nonumber\\
&+&\frac12cb^2[S]S_{\alpha\beta}(A^2 \eta^{\alpha\beta}-2A^\alpha A^\beta ) \nonumber\\
&+&\frac{b^3}{6}([S]^3-3[S][S^2]+2[S^3]) \,,
\end{eqnarray}
which yields the expected cubic vector interactions from above. In particular, the second line is equivalent to $-\frac12a^2b\Ft^{\mu\alpha}\Ft^{\nu}{}_\alpha S_{\mu\nu}$. It is also interesting to notice that the determinantal form also produces the aforementioned couplings through a disformal metric determined by the vector field. Last but not least, the quartic elementary polynomial 
\begin{eqnarray}
e_4 =-\frac{1}{24}\epsilon_{\mu\nu\alpha\beta}\epsilon^{\rho\sigma\delta\gamma}\mathcal{C}^\mu{}_\rho\mathcal{C}^\nu{}_\sigma \mathcal{C}^\alpha{}_\delta \mathcal{C}^\beta{}_\gamma 
\end{eqnarray}
gives the remaining quartic order interactions, which we omit here. Note that the series stop at that order since we are in four dimensions, hence $e_5=0$. 

The determinantal formulation of the vector field interactions also allows to establish an interesting relation with the Born-Infeld action for electromagnetism \cite{BIelectromagnetism}. In the original Born-Infeld theory, the Maxwell lagrangian is replaced by $\sqrt{-\det(\eta_{\mu\nu}+ \lambda^{-2} F_{\mu\nu} )}$ with $\lambda$ some scale. We now notice that our determinant (\ref{determinant}) can be alternatively written as\footnote{Here we encounter the usual problems of the square root of a matrix. To have it well-defined, we assume that $\delta^{\mu}{}_\nu+\mathcal{D}^\mu{}_\nu$ is positive definite.}
\be
\det(\delta^{\mu}{}_\nu+\mathcal{C}^\mu{}_\nu)=\sqrt{\det(\delta^{\mu}{}_\nu+\mathcal{D}^\mu{}_\nu)}
\ee
with $\mathcal{D}^\mu{}_\nu=2\mathcal{C}^\mu{}_\nu+\mathcal{C}^\mu{}_\alpha\mathcal{C}^\alpha{}_\nu$. It is in this sense that our interactions can be regarded as a generalization of Born-Infeld theories to the case of Proca fields. We should emphasize that the pure Born-Infeld lagrangian is very special and we expect many of its properties to be lost in our case, but it would nevertheless be interesting to explore the potential relations.

\section{Curved spacetimes}
In the previous Sections we have explicitly built in a systematic way a finite family of derivative self-interactions for a massive vector field in flat spacetime. We turn now to the case of curved spacetimes and generalise our results in the presence of gravity. As usual, we could follow a {\it minimal coupling principle} and simply replace the partial derivatives in the previous section by covariant derivatives. However, as it is well-known from the case of scalar Galileons, we need to be careful and pay attention not to add new propagating dof's when gravity is turned on. The underlying reason is that our derivative self-interactions lead to dangerous non-minimal couplings that could excite the temporal polarization of the vector field. Fortunately, it is also known that we can add  explicit couplings to the curvature serving as counter-terms to keep the correct number of propagating modes. In the present case we can be guided by the Horndeski interactions, since we know that the Stueckelberg field sector must belong to said family of lagrangians. Since we also have the explicit couplings of the Stueckelberg to the transverse modes, we will have additional terms. With this reasoning in mind, it is not too difficult to obtain the generalisation of the derivative self-interactions (\ref{vecGalProcaField}) to a curved spacetime, which can be written as
\begin{equation}\label{generalizedProfaField_curved}
\mathcal L^{\rm curved}_{\rm gen. Proca} = -\frac14 \sqrt{-g}F_{\mu\nu}^2 + \sqrt{-g}\sum^5_{n=2}\beta_n \mathcal L_n
\end{equation}
where this time the interactions $\mathcal L_n$ become
\begin{eqnarray}\label{vecGalcurv}
\mathcal L_2 & = & G_2(A_\mu,F_{\mu\nu},\Ft_{\mu\nu}) \nonumber\\
\mathcal L_3 & = &G_3(Y)\nabla_\mu A^\mu \nonumber\\
\mathcal L_4 & = & G_{4}(Y)R+G_{4,Y} \left[(\nabla_\mu A^\mu)^2-\nabla_\rho A_\sigma \nabla^\sigma A^\rho\right] \nonumber\\
\mathcal L_5 & = & G_5(Y)G_{\mu\nu}\nabla^\mu A^\nu-\frac{1}{6}G_{5,Y} \Big[
(\nabla\cdot A)^3 \nonumber\\
&+&2\nabla_\rho A_\sigma \nabla^\gamma A^\rho \nabla^\sigma A_\gamma -3(\nabla\cdot A)\nabla_\rho A_\sigma \nabla^\sigma A^\rho \Big] \nonumber \\
&-&\tilde{G}_5(Y) \tilde{F}^{\alpha\mu}\tilde{F}^\beta_{\;\;\mu}\nabla_\alpha A_\beta  \nonumber \\
\mathcal L_6 & = & G_6(Y)L^{\mu\nu\alpha\beta}\nabla_\mu A_\nu \nabla_\alpha A_\beta \nonumber \\
&+&\frac{G_{6,Y}}{2} \tilde{F}^{\alpha\beta}\tilde{F}^{\mu\nu}\nabla_\alpha A_\mu \nabla_\beta A_\nu
\end{eqnarray}
with $\nabla$ denoting the covariant derivative, $Y=-\frac12 A^2$ and $L^{\mu\nu\alpha\beta}$ the double dual Riemann tensor defined as
\begin{eqnarray}
L^{\mu\nu\alpha\beta}=\frac14\epsilon^{\mu\nu\rho\sigma}\epsilon^{\alpha\beta\gamma\delta}R_{\rho\sigma\gamma\delta}.
\end{eqnarray}
A few comments are in order here. As one can see, the requirement of having 3 polarizations for the vector field demands the presence of non-minimal couplings in $\mathcal L_{4,5,6}$. We should however stress that the coupling $\tilde{G}_5(Y) \tilde{F}^{\alpha\mu}\tilde{F}^\beta_{\;\;\mu}\nabla_\alpha A_\beta$ belonging to $\L_5$ does not require any non-minimal counter-term. The reason is the same as for $\L_3$, i.e., that the coupling to the connection is linear. In fact, both terms could be written together as $h^{\mu\nu}\nabla_\mu A_\nu$ with the effective metric $h^{\mu\nu}\equiv G_3g^{\mu\nu}-\tilde{G}_5\Ft ^{\mu\alpha}\Ft^\nu{}_\alpha$. Finally, it is worth emphasizing that in the function $G_2$ we can also include non-minimal couplings of the form 
$G^{\mu\nu}A_\mu A_\nu$, since it does not contain any dynamics for the temporal component of the vector field. In fact, this term can be rewritten via integration by parts as the usual Maxwell term plus a term belonging to $\L_4$ with $G_4=A^2$.

In order to make apparent the interactions of the Stueckelberg field, we will rewrite the above interactions in terms of $F_{\mu\nu}$ and $S_{\mu\nu}$ as
\begin{eqnarray}\label{vecGalcurv}
\mathcal L_2 & = & \hat{G}_2(Y,F,\tilde{F}) \\
\mathcal L_3 & = &\frac12 G_3(Y)[S] \nonumber\\
\mathcal L_4 & = & G_{4}(Y)R+G_{4,Y} \frac{ [S]^2-[S^2]}{4} \nonumber\\
\mathcal L_5 & = & \frac{G_5(Y)}{2}G^{\mu\nu} S_{\mu\nu}-\frac{G_{5,Y}}{6}\frac{ [S]^3-3[S][S^2] +2[S^3]}{8} \nonumber\\
&+&\tilde{G}_5(Y) \tilde{F}^{\mu\alpha}\tilde{F}^\nu_{\;\;\alpha} S_{\mu\nu} \nonumber \\
\mathcal L_6 & = & G_6(Y)L^{\mu\nu\alpha\beta} F_{\mu\nu}F_{\alpha\beta}
+\frac{G_{6,Y}}{2} \tilde{F}^{\alpha\beta}\tilde{F}^{\mu\nu}S_{\alpha\mu} S_{\beta\nu}\nonumber
\end{eqnarray}
where all the additional terms depending only on $F_{\mu\nu}$ have been included in $\L_2$.In order to see that these lagrangians do not propagate additional polarizations for the vector field is useful to keep in mind the relation
\begin{align}
2\nabla_{[\alpha} S_{\beta]\gamma}=&[\nabla_\alpha,\nabla_\beta]A_\gamma+[\nabla_\alpha,\nabla_\gamma]A_\beta-[\nabla_\beta,\nabla_\gamma]A_\alpha\nonumber\\&+\nabla_\gamma F_{\alpha\beta}.
\end{align}
The first line will give couplings to the curvature, while the second term only affects the propagation of the transverse modes. Thus, derivatives of $S$ in the field equations can appear as long as they do it in the above antisymmetric form.

Now it is clear that we recover the Horndeski terms when we replace $S_{\mu\nu}\rightarrow 2\nabla_\mu\nabla_\nu\pi$. However, we see that additional terms survive coupling $\pi$ to $\Ft^{\mu\nu}$. In fact, this coupling induces additional non-minimal couplings in $\L_6$. It is interesting that the counter-term in $\L_6$ is nicely related to the vector-tensor interaction worked out by Horndeski in \cite{Horndeski:1976gi}. This was actually expected since it is known that the only non-minimal coupling for a gauge field is precisely of that form. In fact, Horndeski vector-tensor interaction corresponds to $G_{6,Y}=0$. This particular case has already been studied \cite{Barrow:2012ay} and the consequences of having $G_{6,Y}\neq0$ would be interesting to explore.

%%%%%%%%%%%%%%%%%%%%%%
%%%%%%%%%%%%%%%%%%%%%
\section{Discussion}
This work was devoted to the detailed study of the generalized Proca action with derivative self-interactions keeping 3 polarizations. Starting from the decoupling limit we analyzed which interactions maintain our requirement of absence of Ostrogradski instabilities and motivated how the interactions beyond the decoupling limit can be recovered. We also constructed the allowed interaction terms systematically using the Levi-Civita tensors order by order. Moreover, we showed how the interactions can be generated in a compact way from a determinant. Finally, we generalised our findings to curved spacetime. A very crucial property of the standard scalar Galileon interactions is the non-renormalization theorem, which renders them technically natural. The key ingredient for this property is the construction out of the antisymmetric Levi-Civita tensors. It would be very important to explore the question of whether or not the generalized Proca interactions remain technically natural and push forward the preliminary analysis of  \cite{Charmchi:2015ggf}. We will leave this for a future work.

\acknowledgments We would like to thank Patrick Peter for useful discussions and comments. 
JBJ  acknowledges  the  financial  support  of
A*MIDEX project (n ANR-11-IDEX-0001-02) funded by
the  "Investissements d'Avenir" French Government pro-
gram, managed by the French National Research Agency
(ANR),  MINECO  (Spain) projects
FIS2014-52837-P and Consolider-Ingenio MULTIDARK
CSD2009-00064. L.H. acknowledges financial support from Dr. Max R\"ossler, the Walter Haefner Foundation and the ETH Zurich Foundation.


\begin{thebibliography}{99}

%\cite{Dvali:2000hr}
\bibitem{Dvali:2000hr} 
  G.~R.~Dvali, G.~Gabadadze and M.~Porrati,
  %``4-D gravity on a brane in 5-D Minkowski space,''
  Phys.\ Lett.\ B {\bf 485}, 208 (2000)
  %doi:10.1016/S0370-2693(00)00669-9
  [hep-th/0005016].
  %%CITATION = doi:10.1016/S0370-2693(00)00669-9;%%
  %2187 citations counted in INSPIRE as of 01 Feb 2016

%\cite{Nicolis:2008in}
\bibitem{Nicolis:2008in}
  A.~Nicolis, R.~Rattazzi and E.~Trincherini,
  %``The Galileon as a local modification of gravity,''
  Phys.\ Rev.\ D {\bf 79} (2009) 064036
  %doi:10.1103/PhysRevD.79.064036
  [arXiv:0811.2197 [hep-th]].
  %%CITATION = doi:10.1103/PhysRevD.79.064036;%%
  %786 citations counted in INSPIRE as of 31 janv. 2016
  
  %\cite{quantum_corrections}
\bibitem{quantum_corrections} 
  K.~Hinterbichler, M.~Trodden and D.~Wesley,
  %``Multi-field galileons and higher co-dimension branes,''
  Phys.\ Rev.\ D {\bf 82}, 124018 (2010)
  %doi:10.1103/PhysRevD.82.124018
  [arXiv:1008.1305 [hep-th]];
  %%CITATION = %doi:10.1103/PhysRevD.82.124018;%%
  %140 citations counted in INSPIRE as of 08 févr. 2016
  %\cite{Brouzakis:2013lla}
%\bibitem{Brouzakis:2013lla} 
  N.~Brouzakis, A.~Codello, N.~Tetradis and O.~Zanusso,
  %``Quantum corrections in Galileon theories,''
  Phys.\ Rev.\ D {\bf 89}, no. 12, 125017 (2014)
  %doi:10.1103/PhysRevD.89.125017
  [arXiv:1310.0187 [hep-th]];
  %%CITATION = doi:10.1103/PhysRevD.89.125017;%%
  %13 citations counted in INSPIRE as of 08 Feb 2016
%\cite{dePaulaNetto:2012hm}
%\bibitem{dePaulaNetto:2012hm} 
  T.~de Paula Netto and I.~L.~Shapiro,
  %``One-loop divergences in the Galileon model,''
  Phys.\ Lett.\ B {\bf 716}, 454 (2012)
  %doi:10.1016/j.physletb.2012.08.056
  [arXiv:1207.0534 [hep-th]].
  %%CITATION = doi:10.1016/j.physletb.2012.08.056;%%
  %13 citations counted in INSPIRE as of 08 Feb 2016
%\cite{Heisenberg:2014raa}
%\bibitem{Heisenberg:2014raa} 
  L.~Heisenberg,
  %``Quantum Corrections in Galileons from Matter Loops,''
  Phys.\ Rev.\ D {\bf 90}, no. 6, 064005 (2014)
  %doi:10.1103/PhysRevD.90.064005
  [arXiv:1408.0267 [hep-th]];
  %%CITATION = doi:10.1103/PhysRevD.90.064005;%%
  %9 citations counted in INSPIRE as of 08 Feb 2016
%\cite{deRham:2012ew}
%\bibitem{deRham:2012ew} 
  C.~de Rham, G.~Gabadadze, L.~Heisenberg and D.~Pirtskhalava,
  %``Nonrenormalization and naturalness in a class of scalar-tensor theories,''
  Phys.\ Rev.\ D {\bf 87}, no. 8, 085017 (2013)
  %doi:10.1103/PhysRevD.87.085017
  [arXiv:1212.4128].
  %%CITATION = doi:10.1103/PhysRevD.87.085017;%%
  %64 citations counted in INSPIRE as of 08 Feb 2016


  
  
  %\cite{deRham:2010ik}
\bibitem{deRham:2010ik} 
  C.~de Rham and G.~Gabadadze,
  %``Generalization of the Fierz-Pauli Action,''
  Phys.\ Rev.\ D {\bf 82}, 044020 (2010)
  %doi:10.1103/PhysRevD.82.044020
  [arXiv:1007.0443 [hep-th]].
  %%CITATION = doi:10.1103/PhysRevD.82.044020;%%
  %533 citations counted in INSPIRE as of 01 fŽvr. 2016
  %\cite{deRham:2010tw}
\bibitem{deRham:2010tw} 
  C.~de Rham, G.~Gabadadze, L.~Heisenberg and D.~Pirtskhalava,
  %``Cosmic Acceleration and the Helicity-0 Graviton,''
  Phys.\ Rev.\ D {\bf 83}, 103516 (2011)
  %doi:10.1103/PhysRevD.83.103516
  [arXiv:1010.1780 [hep-th]];
  %%CITATION = doi:10.1103/PhysRevD.83.103516;%%
  %142 citations counted in INSPIRE as of 01 fŽvr. 2016
 %\cite{deRham:2010kj}
%\bibitem{deRham:2010kj} 
  C.~de Rham, G.~Gabadadze and A.~J.~Tolley,
  %``Resummation of Massive Gravity,''
  Phys.\ Rev.\ Lett.\  {\bf 106}, 231101 (2011)
  %doi:10.1103/PhysRevLett.106.231101
  [arXiv:1011.1232 [hep-th]];
  %%CITATION = doi:10.1103/PhysRevLett.106.231101;%%
  %667 citations counted in INSPIRE as of 01 Feb 2016
  %\cite{deRham:2011rn}
%\bibitem{deRham:2011rn} 
  C.~de Rham, G.~Gabadadze and A.~J.~Tolley,
  %``Ghost free Massive Gravity in the St\'uckelberg language,''
  Phys.\ Lett.\ B {\bf 711}, 190 (2012)
  %doi:10.1016/j.physletb.2012.03.081
  [arXiv:1107.3820 [hep-th]];
  %%CITATION = doi:10.1016/j.physletb.2012.03.081;%%
  %174 citations counted in INSPIRE as of 02 fŽvr. 2016
%\cite{Hassan:2012qv}
%\bibitem{Hassan:2012qv} 
  S.~F.~Hassan, A.~Schmidt-May and M.~von Strauss,
  %``Proof of Consistency of Nonlinear Massive Gravity in the St\'uckelberg Formulation,''
  Phys.\ Lett.\ B {\bf 715}, 335 (2012)
 % doi:10.1016/j.physletb.2012.07.018
  [arXiv:1203.5283 [hep-th]].
  %%CITATION = doi:10.1016/j.physletb.2012.07.018;%%
  %110 citations counted in INSPIRE as of 08 Feb 2016


%\cite{Vainshtein:1972sx}
\bibitem{Vainshtein:1972sx} 
  A.~I.~Vainshtein,
  %``To the problem of nonvanishing gravitation mass,''
  Phys.\ Lett.\ B {\bf 39}, 393 (1972).
  %doi:10.1016/0370-2693(72)90147-5
  %%CITATION = doi:10.1016/0370-2693(72)90147-5;%%
  %752 citations counted in INSPIRE as of 08 févr. 2016



%\cite{Babichev:2011iz}
\bibitem{Babichev:2011iz}
  E.~Babichev, C.~Deffayet and G.~Esposito-Farese,
  %``Constraints on Shift-Symmetric Scalar-Tensor Theories with a Vainshtein Mechanism from Bounds on the Time Variation of G,''
  Phys.\ Rev.\ Lett.\  {\bf 107} (2011) 251102
 % doi:10.1103/PhysRevLett.107.251102
  [arXiv:1107.1569 [gr-qc]].
  %%CITATION = doi:10.1103/PhysRevLett.107.251102;%%

%\cite{Jimenez:2015bwa}
\bibitem{Jimenez:2015bwa}
  J.~Beltr\'an Jim\'enez, F.~Piazza and H.~Velten,
  %``Piercing the Vainshtein screen with anomalous gravitational wave speed: Constraints on modified gravity from binary pulsars,''
  Phys.\ Rev.\ Lett. {\bf 116} (2016) 061101 
  arXiv:1507.05047 [gr-qc].
  %%CITATION = ARXIV:1507.05047;%%
  %11 citations counted in INSPIRE as of 09 Feb 2016



%\cite{Horndeski:1974wa}
\bibitem{Horndeski:1974wa} 
  G.~W.~Horndeski,
  %``Second-order scalar-tensor field equations in a four-dimensional space,''
  Int.\ J.\ Theor.\ Phys.\  {\bf 10}, 363 (1974).
%  doi:10.1007/BF01807638;
  %%CITATION = doi:10.1007/BF01807638;%%
  %434 citations counted in INSPIRE as of 08 févr. 2016
  %\cite{Deffayet:2011gz}

%\bibitem{Deffayet:2009mn}
  C.~Deffayet, S.~Deser and G.~Esposito-Farese,
  %``Generalized Galileons: All scalar models whose curved background extensions maintain second-order field equations and stress-tensors,''
  Phys.\ Rev.\ D {\bf 80} (2009) 064015
  %doi:10.1103/PhysRevD.80.064015
  [arXiv:0906.1967 [gr-qc]].
  %%CITATION = doi:10.1103/PhysRevD.80.064015;%%
  %327 citations counted in INSPIRE as of 09 fŽvr. 2016

%\bibitem{Deffayet:2011gz} 
  C.~Deffayet, X.~Gao, D.~A.~Steer and G.~Zahariade,
  %``From k-essence to generalised Galileons,''
  Phys.\ Rev.\ D {\bf 84}, 064039 (2011)
  %doi:10.1103/PhysRevD.84.064039
  [arXiv:1103.3260 [hep-th]].
  %%CITATION = doi:10.1103/PhysRevD.84.064039;%%
  %316 citations counted in INSPIRE as of 08 févr. 2016
  %\cite{Deffayet:2009mn}

%\cite{deRham:2011by}
\bibitem{deRham:2011by} 
  C.~de Rham and L.~Heisenberg,
  %``Cosmology of the Galileon from Massive Gravity,''
  Phys.\ Rev.\ D {\bf 84}, 043503 (2011)
  %doi:10.1103/PhysRevD.84.043503
  [arXiv:1106.3312 [hep-th]].
  %%CITATION = doi:10.1103/PhysRevD.84.043503;%%
  %95 citations counted in INSPIRE as of 08 Feb 2016

\bibitem{beyondH}
%\cite{Zumalacarregui:2013pma}
%\bibitem{Zumalacarregui:2013pma} 
  M.~Zumalac\'arregui and J.~Garc\'ia-Bellido,
  %``Transforming gravity: from derivative couplings to matter to second-order scalar-tensor theories beyond the Horndeski Lagrangian,''
  Phys.\ Rev.\ D {\bf 89}, 064046 (2014)
  %doi:10.1103/PhysRevD.89.064046
  [arXiv:1308.4685 [gr-qc]].
%%CITATION = doi:10.1103/PhysRevD.89.064046;%%
  %75 citations counted in INSPIRE as of 08 Feb 2016
%\cite{Gleyzes:2014dya}
%\bibitem{Gleyzes:2014dya} 
  J.~Gleyzes, D.~Langlois, F.~Piazza and F.~Vernizzi,
  %``Healthy theories beyond Horndeski,''
  Phys.\ Rev.\ Lett.\  {\bf 114}, no. 21, 211101 (2015)
  %doi:10.1103/PhysRevLett.114.211101
  [arXiv:1404.6495 [hep-th]];
  %%CITATION = doi:10.1103/PhysRevLett.114.211101;%%
  %91 citations counted in INSPIRE as of 08 févr. 2016
  
%\cite{Langlois:2015cwa}
%\bibitem{Langlois:2015cwa}
  D.~Langlois and K.~Noui,
  %``Degenerate higher derivative theories beyond Horndeski: evading the Ostrogradski instability,''
  arXiv:1510.06930 [gr-qc].
  %%CITATION = ARXIV:1510.06930;%%
  %9 citations counted in INSPIRE as of 09 fŽvr. 2016

%\cite{Langlois:2015skt}
%\bibitem{Langlois:2015skt}
  D.~Langlois and K.~Noui,
  %``Hamiltonian analysis of higher derivative scalar-tensor theories,''
  arXiv:1512.06820 [gr-qc].
  %%CITATION = ARXIV:1512.06820;%%
  %2 citations counted in INSPIRE as of 09 Feb 2016


%\cite{deRham:2010eu}
\bibitem{deRham:2010eu} 
  C.~de Rham and A.~J.~Tolley,
  %``DBI and the Galileon reunited,''
  JCAP {\bf 1005}, 015 (2010)
  %doi:10.1088/1475-7516/2010/05/015
  [arXiv:1003.5917 [hep-th]].
  %%CITATION = doi:10.1088/1475-7516/2010/05/015;%%
  %214 citations counted in INSPIRE as of 08 févr. 2016
  
  %\cite{VanAcoleyen:2011mj}
%\bibitem{VanAcoleyen:2011mj}
  K.~Van Acoleyen and J.~Van Doorsselaere,
  %``Galileons from Lovelock actions,''
  Phys.\ Rev.\ D {\bf 83} (2011) 084025
  %doi:10.1103/PhysRevD.83.084025
  [arXiv:1102.0487 [gr-qc]].
  %%CITATION = doi:10.1103/PhysRevD.83.084025;%%
  %68 citations counted in INSPIRE as of 09 Feb 2016

  
  %\cite{Deffayet:2010zh}
\bibitem{Deffayet:2010zh}
  C.~Deffayet, S.~Deser and G.~Esposito-Farese,
  %``Arbitrary $p$-form Galileons,''
  Phys.\ Rev.\ D {\bf 82} (2010) 061501
  %doi:10.1103/PhysRevD.82.061501
  [arXiv:1007.5278 [gr-qc]].
  %%CITATION = doi:10.1103/PhysRevD.82.061501;%%
  %119 citations counted in INSPIRE as of 31 janv. 2016

%\cite{Deffayet:2016von}
\bibitem{Deffayet:2016von}
  C.~Deffayet, S.~Mukohyama and V.~Sivanesan,
  %``On p-form theories with gauge invariant second order field equations,''
  arXiv:1601.01287 [hep-th].
  %%CITATION = ARXIV:1601.01287;%%
   
   
     %\cite{Jimenez:2014rna}
\bibitem{Jimenez:2014rna}
  J.~Beltran Jimenez and T.~S.~Koivisto,
  %``Extended Gauss-Bonnet gravities in Weyl geometry,''
  Class.\ Quant.\ Grav.\  {\bf 31} (2014) 135002
  %doi:10.1088/0264-9381/31/13/135002
  [arXiv:1402.1846 [gr-qc]].
  %%CITATION = doi:10.1088/0264-9381/31/13/135002;%%
  %17 citations counted in INSPIRE as of 31 Jan 2016

%\cite{Jimenez:2015fva}
\bibitem{Jimenez:2015fva}
  J.~Beltran Jimenez and T.~S.~Koivisto,
  %``Spacetimes with vector distortion: Inflation from generalised Weyl geometry,''
  arXiv:1509.02476 [gr-qc].
  %%CITATION = ARXIV:1509.02476;%%
  %2 citations counted in INSPIRE as of 31 janv. 2016  

   
   
   
  
  %\cite{Heisenberg:2014rta}
\bibitem{Heisenberg:2014rta}
  L.~Heisenberg,
  %``Generalization of the Proca Action,''
  JCAP {\bf 1405} (2014) 015
  %doi:10.1088/1475-7516/2014/05/015
  [arXiv:1402.7026 [hep-th]].
  %%CITATION = doi:10.1088/1475-7516/2014/05/015;%%
  %29 citations counted in INSPIRE as of 31 janv. 2016
  
  
  %\cite{BeltranJimenez:2010uh}
\bibitem{BeltranJimenez:2010uh}
  J.~Beltran Jimenez and A.~L.~Maroto,
  %``Dark energy, non-minimal couplings and the origin of cosmic magnetic fields,''
  JCAP {\bf 1012} (2010) 025
  %doi:10.1088/1475-7516/2010/12/025
  [arXiv:1010.4513 [astro-ph.CO]].
  %%CITATION = doi:10.1088/1475-7516/2010/12/025;%%
  %14 citations counted in INSPIRE as of 31 Jan 2016
  %\cite{Jimenez:2010hu}

    
  
  %\cite{Tasinato:2014eka}
\bibitem{Tasinato:2014eka}
  G.~Tasinato,
  %``Cosmic Acceleration from Abelian Symmetry Breaking,''
  JHEP {\bf 1404} (2014) 067
  %doi:10.1007/JHEP04(2014)067
  [arXiv:1402.6450 [hep-th]];
  %%CITATION = doi:10.1007/JHEP04(2014)067;%%
  %29 citations counted in INSPIRE as of 31 Jan 2016
  
  %\cite{Tasinato:2014mia}
%\bibitem{Tasinato:2014mia}
  G.~Tasinato,
  %``A small cosmological constant from Abelian symmetry breaking,''
  Class.\ Quant.\ Grav.\  {\bf 31} (2014) 225004
  %doi:10.1088/0264-9381/31/22/225004
  [arXiv:1404.4883 [hep-th]].
  %%CITATION = doi:10.1088/0264-9381/31/22/225004;%%
  %7 citations counted in INSPIRE as of 31 janv. 2016
  %\cite{Hull:2015uwa}
\bibitem{Hull:2015uwa}
  M.~Hull, K.~Koyama and G.~Tasinato,
  %``Covariantised Vector Galileons,''
  arXiv:1510.07029 [hep-th].
  %%CITATION = ARXIV:1510.07029;%%
  %4 citations counted in INSPIRE as of 31 Jan 2016


%\cite{Allys:2015sht}
\bibitem{Allys:2015sht}
  E.~Allys, P.~Peter and Y.~Rodriguez,
  %``Generalized Proca action for an Abelian vector field,''
  arXiv:1511.03101 [hep-th].
  %%CITATION = ARXIV:1511.03101;%%
  %2 citations counted in INSPIRE as of 31 janv. 2016

%\cite{Boulware:1973my}
\bibitem{Boulware:1973my} 
  D.~G.~Boulware and S.~Deser,
  %``Can gravitation have a finite range?,''
  Phys.\ Rev.\ D {\bf 6}, 3368 (1972).
  %doi:10.1103/PhysRevD.6.3368
  %%CITATION = doi:10.1103/PhysRevD.6.3368;%%
  %623 citations counted in INSPIRE as of 08 Feb 2016

%\cite{Zhou:2011ix}
\bibitem{Zhou:2011ix}
  S.~Y.~Zhou and E.~J.~Copeland,
  %``Galileons with Gauge Symmetries,''
  Phys.\ Rev.\ D {\bf 85} (2012) 065002
  %doi:10.1103/PhysRevD.85.065002
  [arXiv:1112.0968 [hep-th]].
  %%CITATION = doi:10.1103/PhysRevD.85.065002;%%
  %19 citations counted in INSPIRE as of 15 fŽvr. 2016
%\cite{Goon:2012mu}
\bibitem{Goon:2012mu}
  G.~Goon, K.~Hinterbichler, A.~Joyce and M.~Trodden,
  %``Gauged Galileons From Branes,''
  Phys.\ Lett.\ B {\bf 714} (2012) 115
 % doi:10.1016/j.physletb.2012.06.065
  [arXiv:1201.0015 [hep-th]].
  %%CITATION = doi:10.1016/j.physletb.2012.06.065;%%
  %16 citations counted in INSPIRE as of 15 fŽvr. 2016


%\cite{Tasinato:2013oja}
\bibitem{Tasinato:2013oja}
  G.~Tasinato, K.~Koyama and N.~Khosravi,
  %``The role of vector fields in modified gravity scenarios,''
  JCAP {\bf 1311} (2013) 037
  %doi:10.1088/1475-7516/2013/11/037
  [arXiv:1307.0077 [hep-th]].
  %%CITATION = doi:10.1088/1475-7516/2013/11/037;%%
  %6 citations counted in INSPIRE as of 01 Feb 2016
  

%\cite{deRham:2014lqa}
\bibitem{deRham:2014lqa}
  C.~De Rham, L.~Keltner and A.~J.~Tolley,
  %``Generalized galileon duality,''
  Phys.\ Rev.\ D {\bf 90} (2014) 2,  024050
 % doi:10.1103/PhysRevD.90.024050
  [arXiv:1403.3690 [hep-th]].
  %%CITATION = doi:10.1103/PhysRevD.90.024050;%%
  %28 citations counted in INSPIRE as of 18 Feb 2016


\bibitem{BIelectromagnetism}
%\cite{Born:1934gh}
%\bibitem{Born:1934gh}
  M.~Born and L.~Infeld,
  %``Foundations of the new field theory,''
  Proc.\ Roy.\ Soc.\ Lond.\ A {\bf 144} (1934) 425.
  %doi:10.1098/rspa.1934.0059
  %%CITATION = doi:10.1098/rspa.1934.0059;%%
  %795 citations counted in INSPIRE as of 18 Feb 2016
  %\cite{Ketov:2001dq}
%\bibitem{Ketov:2001dq} 
  S.~V.~Ketov,
  %``Many faces of Born-Infeld theory,''
  hep-th/0108189.
  %%CITATION = HEP-TH/0108189;%%
  %24 citations counted in INSPIRE as of 15 févr. 2016


%\cite{Horndeski:1976gi}
\bibitem{Horndeski:1976gi}
  G.~W.~Horndeski,
  %``Conservation of Charge and the Einstein-Maxwell Field Equations,''
  J.\ Math.\ Phys.\  {\bf 17} (1976) 1980.
  %doi:10.1063/1.522837
  %%CITATION = doi:10.1063/1.522837;%%
  %50 citations counted in INSPIRE as of 18 Feb 2016

  
%\cite{Barrow:2012ay}
\bibitem{Barrow:2012ay}
  J.~D.~Barrow, M.~Thorsrud and K.~Yamamoto,
  %``Cosmologies in Horndeski's second-order vector-tensor theory,''
  JHEP {\bf 1302} (2013) 146
  %doi:10.1007/JHEP02(2013)146
  [arXiv:1211.5403 [gr-qc]].
  %%CITATION = doi:10.1007/JHEP02(2013)146;%%
  %13 citations counted in INSPIRE as of 18 Feb 2016
%\cite{Jimenez:2013qsa}
%\bibitem{Jimenez:2013qsa}
  J.~Beltran Jimenez, R.~Durrer, L.~Heisenberg and M.~Thorsrud,
  %``Stability of Horndeski vector-tensor interactions,''
  JCAP {\bf 1310} (2013) 064
  %doi:10.1088/1475-7516/2013/10/064
  [arXiv:1308.1867 [hep-th]].
  %%CITATION = doi:10.1088/1475-7516/2013/10/064;%%
  %30 citations counted in INSPIRE as of 18 Feb 2016
  
  
  
  %\cite{Charmchi:2015ggf}
\bibitem{Charmchi:2015ggf} 
  F.~Charmchi, Z.~Haghani, S.~Shahidi and L.~Shahkarami,
  %``One-Loop Corrections to Vector Galileon Theory,''
  arXiv:1511.07034 [hep-th].
  %%CITATION = ARXIV:1511.07034;%%
  %1 citations counted in INSPIRE as of 09 Feb 2016



\end{thebibliography}
\end{document}